# Electronic nematicity in the absence of charge density waves in a new titanium-based kagome metal

Hong Li[1], Siyu Cheng[1], Brenden R. Ortiz[2], Hengxin Tan[3], Dominik Werhahn[4], Keyu Zeng[1], Dirk Johrendt[4], Binghai Yan[3], Ziqiang Wang[1], Stephen D. Wilson[2] and Ilija Zeljkovic[1]

[1] Department of Physics, Boston College, Chestnut Hill, MA 02467, USA
[2] Materials Department, University of California Santa Barbara, Santa Barbara, California 93106, USA
[3] Department of Condensed Matter Physics, Weizmann Institute of Science, Rehovot, Israel
[4] Department of Chemistry, Ludwig-Maximilians-Universität München, 81377 München, Germany

**Layered crystalline materials that consist of transition metal atoms laid out on a kagome network emerged as a versatile platform to study novel electronic phenomena. Within this realm, vanadium-based kagome superconductors $AV_3Sb_5$ ($A$=K, Cs, Rb) recently attracted a large interest for various exotic electronic states. These states arise within a "smectic" parent charge density wave (CDW) phase that appears to simultaneously break both the translational and the rotational symmetry of the lattice. We use spectroscopic-imaging scanning tunneling microscopy to study a newly synthesized kagome metal $CsTi_3Bi_5$, which is isostructural to $AV_3Sb_5$ but with a titanium atom kagome network. In sharp contrast to its vanadium-based counterpart, $CsTi_3Bi_5$ does not exhibit detectable CDW states. By imaging the scattering and interference of electrons, we obtain momentum-space information about the low-energy electronic structure of $CsTi_3Bi_5$. A comparison to our density functional theory calculations reveals substantial electronic correlation effects at low-energies in this material. Remarkably, by comparing the amplitudes of scattering wave vectors along different directions, we discover a pronounced electronic anisotropy that breaks the six-fold symmetry of the lattice, arising from both in-plane and out-of-plane titanium-derived *d* orbitals. Our experiments reveal the emergence of a genuine electronic nematic order in $CsTi_3Bi_5$, the first among the "135" family of kagome metals, which is in contrast to the vanadium-based $AV_3Sb_5$ where rotation symmetry breaking appears to be intertwined with the CDW. Our work also uncovers the significant role of electronic orbitals in the kagome metal $CsTi_3Bi_5$, suggestive of a hexagonal analogue of the nematic bond order in Fe-based superconductors.**

Quantum materials composed of atoms arranged on a kagome net of corner-sharing triangles present an exciting platform to realize novel electronic behavior. Out of the wide array of transition metal atoms that can be used to populate the kagome layers in materials synthesized thus far, materials based on vanadium kagome layers in the $AV_3Sb_5$ ($A$=Cs, K, Rb) structure emerged as rare example of a kagome superconductor [1–4], which triggered an avalanche of experimental [5–18] and theoretical [19–26] work. Experiments demonstrated that $AV_3Sb_5$ indeed exhibits the canonical features of the band structure naturally arising due to the intrinsic geometric frustration of the kagome lattice, including Dirac fermions, flat bands and multiple van Hove singularities [14,27–29]. At the same time, the system showcased intriguing similarities to correlated electron phenomena observed in high-temperature superconductors, such as charge density waves (CDWs) [5,6,11,12], a pair density wave [11] and electronic directionality that breaks the $C_6$ rotation symmetry of the lattice [5–7,11,15,30,31].

Rotation symmetry breaking in particular has been studied in connection to the three-dimensional CDW state in this family of materials, which is the first ordered phase that emerges upon cooling down the system below $T^* \approx$ 80-100 K [2,4,13,18,32,33]. Mounting experimental evidence suggests that rotation symmetry breaking may be intertwined with the emergence of the CDW state – it occurs concomitant with the formation of the

CDW [34,35] and it has been reported that rotation symmetry breaking is CDW-driven [30]. Theoretically, rotation symmetry breaking in $AV_3Sb_5$ could be explained by the inter-layer stacking order of the CDW phase, which is intrinsically three-dimensional and in turn selects one out of the three lattice directions as the preferred one [23,24]. While other distinct experimental signatures of rotation symmetry breaking are found to onset at an even lower temperatures, for example the charge-stripe order [6], the anisotropy in elasto-transport measurements [30] and unidirectional coherent quasiparticles [35], these all inevitably occur within the system where both rotation and translation symmetries appear to be already broken at the CDW onset $T^*$ [34,35]. As such, within the liquid crystal classification of electronic phases, the parent electronic order of $AV_3Sb_5$ could be regarded as a "smectic" phase. This brings a fundamental question – to what extent is rotation symmetry breaking in kagome metals tied to the CDW order itself and can it exist in the absence of translation symmetry breaking to form a genuine electronic "nematic" phase.

Recently a new family of Ti-based kagome metals $ATi_3Bi_5$ ($A$=Cs, Rb) has been synthesized [36,37] in the same crystal structure as $AV_3Sb_5$, but with a kagome net of Ti atoms replacing the V, and Bi substituting for Sb. While some studies suggested that $CsTi_3Bi_5$ may be superconducting below critical temperature $T_c$~4 K [36] similarly to $CsV_3Sb_5$, superconductivity has not been confirmed by other studies yet [37]. From the fermiology perspective, $ATi_3Bi_5$ also hosts van Hove singularities at M points, but these are now positioned well above the Fermi level [38] unlike the equivalent features in $AV_3Sb_5$ that appear in closer proximity to the Fermi level [14]. In contrast to $AV_3Sb_5$, experiments on $ATi_3Bi_5$ reveal no obvious anomalies in magnetization, resistivity and heat capacity [37] that could be attributed to a CDW state, which has been widely observed in $AV_3Sb_5$. This in turn provides an opportunity to explore novel electronic phenomena in a non-magnetic kagome metal in the absence of the translation symmetry breaking CDW. Our scanning tunneling microscopy measurements of cleaved $CsTi_3Bi_5$ bulk single crystals are consistent with the absence of a CDW in this system. Moreover, we use spectroscopic-imaging scanning tunneling microscopy (SI-STM) to discover a pronounced rotation symmetry breaking in the electronic signal. In particular, we observe substantial directionality in the amplitudes of scattering wave vectors between the three nominally identical high-symmetry directions. We find that anisotropic spectral weight associated with different portions of the bands can naturally explain unidirectionality in our SI-STM measurements. In sharp contrast to $AV_3Sb_5$ where rotation symmetry breaking signatures occur within the CDW phase that already breaks the rotation symmetry, the reduction of the rotation symmetry here clearly occurs in the absence of the concomitant translation symmetry breaking, thus forming a pure electronic nematic phase. This is akin to the electronic nematic phase in Fe-based high-temperature superconductors [39–44], but on a hexagonal lattice.

Bulk single crystals of $CsTi_3Bi_5$ exhibit a hexagonal crystal structure ($a \approx 5.7$ Å, $c \approx 9.2$ Å) [37] composed of Ti-Bi slabs stacked between alkali Cs layers (Fig. 1a,b). The Ti kagome net is interlaced with Bi atoms, and sandwiched between two other hexagonal Bi layers (Fig. 1a). Due to the air sensitive nature of the samples, we handle sample preparation in an inert environment of an argon glove box and minimize subsequent exposure to air during the transfer to the STM chamber (Methods). The samples are cleaved in UHV at cryogenic temperatures and immediately inserted into the STM for measurements. Similarly to the cleaving structure extensively explored in $AV_3Sb_5$, $CsTi_3Bi_5$ crystals should naturally cleave between the Cs and the Bi layers, resulting in two different surface terminations. Indeed, STM topographs reveal the two main types of atomically-ordered surface morphologies: an incomplete layer that we attribute to the Cs layer on top of a complete honeycomb-like surface structure, which we ascribe to the Bi layer (Fig. 1c,d). While STM topographs of both surfaces can show atomic resolution (Fig. 1e,f), we note that the Cs surface tends to form disconnected island-like patches and that large complete regions of the Cs surface are difficult to locate. In contrast, Bi-terminated layers are robust, and we can find large flat areas for imaging (Fig. 2a).

Fourier transforms of low-temperature STM topographs of the Bi surface show a hexagonal lattice with wave vectors $a_{STM} \approx b_{STM} \approx 5.8$ Å (inset in Fig. 2a), which are consistent with the expected bulk lattice constants. Aside from atomic Bragg peaks $\mathbf{Q}_{Bragg}$, no additional peaks are detected along high-symmetry directions in the Fourier transforms (FTs) of STM topographs (Fig. 2b). This is consistent with the absence of a CDW in $CsTi_3Bi_5$, in contrast to the CDW widely observed in STM topographs of the $AV_3Sb_5$ family acquired over a wide range of biases [5–7,11,12,45].

To gain insight into the electronic band structure of $CsTi_3Bi_5$, we perform SI-STM measurements, which are rooted in elastic scattering and interference of electrons detectable as periodic modulations in differential conductance $dI/dV(\mathbf{r}, V)$ maps. Quantitative assessment of different scattering wave vectors can be more easily obtained from discrete FTs of $dI/dV(\mathbf{r},V)$ maps, which provide momentum-space information. The FTs of $dI/dV(\mathbf{r},V)$ maps acquired at high energies on the Bi surface of $CsTi_3Bi_5$ show a series of dispersive scattering wave vectors (Fig. 2c,d). To understand their origin, we focus on the three concentric wave vectors that appear over a wide energy range, labeled $\mathbf{q}_1$, $\mathbf{q}_2$ and $\mathbf{q}_3$ (Fig. 2d,e,g). To visualize the dispersion of these vectors, we first extract radially-averaged FT linecuts of normalized $dI/dV$ maps as a function of energy (Fig. 2e). We also extract a linear FT linecut to determine the Fermi velocities of different bands (Fig. 2g). We find that they exhibit large and positive dispersion velocities: 5200±230 meV·Å for $\mathbf{q}_1$, 3330±70 meV·Å for $\mathbf{q}_2$ and 2370±100 meV·Å for $\mathbf{q}_3$ (Supplementary Figure 1). We compare the morphology and the dispersion of these wave vectors in SI-STM data to density functional theory (DFT) calculations of $CsTi_3Bi_5$ including spin-orbit coupling (Fig. 2f, Methods). DFT calculations show several electron-like bands centered at Γ. We attribute the inner $\mathbf{q}_1$ wave vector to the intra-band scattering within the Bi-derived $p$ orbital pocket at Γ due to a close agreement in dispersion velocities, Fermi wave vectors and the band bottom energies between theory and experiment (Fig. 2e-h). The outer wave vectors, $\mathbf{q}_2$ and $\mathbf{q}_3$, arise from scattering between Ti $d_{xy}/d_{x2-y2}$ band and Bi $p$ band ($\mathbf{q}_2$) and intra-band scattering within the Ti $d_{xy}/d_{x2-y2}$ band ($\mathbf{q}_3$). We can conclude this based on the morphology of the wave vectors and comparable dispersion velocities (Fig. 2g,h). Interestingly however, there is a sizeable difference in the energy of Ti-derived bands compared to what is expected from theory (Fig. 2h), suggesting a relative energy offset between the Bi and Ti derived bands not captured by theory. We note that this cannot be explained by a rigid band shift of all bands induced by small surface doping. While an unusual band-selective surface doping could contribute to this behavior, we deem this to be unlikely as relevant portions of bands involved in scattering in our STM data appear to be pushed down by electron doping in a comparable manner [38]. This in turn may already hint at additional Coulomb interaction and correlation effects in this material.

Visual inspection of the FTs leads us to another intriguing observation – while each wave vector should appear to be the same along nominally equivalent high-symmetry directions, some of them exhibit a noticeable intensity anisotropy (Fig. 3a,d). In particular, $\mathbf{q}_2$ and $\mathbf{q}_3$ along one direction (purple dashed line) appear to be more intense compare to the other two nominally equivalent directions. To quantify the angle-dependent amplitude variation, we plot the FT amplitudes of the peaks as a function of angle (Fig. 3b,c). Both show the same dominant axis. To confirm the robustness of this observation, we evaluate potential tip anisotropy artifacts that could cause the apparent FT intensity anisotropy. We first note that we repeated the experiment on multiple samples with different tip wires (Fig. 3a,g,h, Supplementary Figure 3). We then show that the intensity anisotropy is robust to rotating the scan direction by 90 degrees (Supplementary Figure 2). We stress that we have changed the microscopic arrangement of atoms at the tip apex *in-situ* by fast scanning and bias pulsing multiple times, and have still consistently observed the same symmetry axis over the same area of the sample (Supplementary Figure 3). In addition to the intensity anisotropy of the scattering wave vectors, we also find angular dependence of the scattering wave vector magnitudes (Fig. 3e,f). While this angular variation of vector magnitudes is only at the order of a few percent, it directly demonstrates small anisotropy of the associated constant energy contour, elongated along the same direction as the symmetry axis observed in vector intensity plots in Fig. 3a. We note that we have not been able to locate a domain wall across which the dominant axis would rotate by 120 degrees, possibly due to a larger domain size compared to the CDW domains in $AV_3Sb_5$ where we have occasionally achieved this [6,7]. However, in place of this test, we proceed to map the apparent symmetry axis before and after a manual rotation of the sample by 120° while keeping the same orientation of the tip wire (Fig. 3g-j). If breaking of rotational symmetry observed in Figure 3 was an artifact due to the tip anisotropy, we would expect to still observe the same orientation of a tip artifact induced $C_2$-symmetry axis in equivalent SI-STM scans. Importantly however, the orientation of the symmetry axis is now also rotated by 120° (Fig. 3g compared to Fig. 3h), pointing against a spurious $C_2$-symmetric signal caused by tip anisotropy. Taken together, our experiments reveal that the rotation symmetry of the electronic structure of $CsTi_3Bi_5$ is reduced from $C_6$-symmetric to $C_2$-symmetric.

We proceed to examine the orbital structure of electronic states associated with scattering wave vectors in more detail, by examining constant energy contours (CECs) obtained from our DFT calculations. The Fermi surface consists of several closed contours (inset in Fig. 4b), which show a weak $k_z$-dependence (Supplementary Figure 4). The small circular pocket centered at $\Gamma$ is primarily composed of Bi $p_z$ orbitals, while the rest of the bands are largely derived from different Ti $d$ orbitals. For pedagogical purposes, we illustrate what subsets of bands in the CEC play a role in the emergence of each set of features seen in the experimental FT (top panels in Fig. 4c-f), and then perform auto-correlation of the CEC to mimic the features in the experimental FT (bottom row in Fig. 4c-f). For example, as shown in Fig. 4c, we find that the previously discussed wave vectors in Fig. 2 can be captured by scattering between: $p_z$ orbitals ($\mathbf{q}_1$), $p_z$ and $d_{xy}/d_{x2-y2}$ orbitals ($\mathbf{q}_2$) and $d_{xy}/d_{x2-y2}$ orbitals ($\mathbf{q}_3$). In addition to these previously discussed features, FTs at low energies also show a series of elongated quasi-1D features near the Brillouin zone edge (black rectangle in Fig. 4a). These features clearly disperse with energy (Fig. 4h,i) and can be explained by an interplay of scattering and interference of electrons primarily from $d_{xy}/d_{x2-y2}$ and $d_{xz}/d_{yz}$ orbitals (Fig. 4d-f). These elongated scattering wave vectors near the Brillouin zone edge appear more intense along one direction (black rectangle in Fig. 4a), and look tantalizingly similar to the quasi-1D vectors in $CsV_3Sb_5$ [6,11,35] that also arise from $d$ orbital scattering. Putting together the aforementioned scattering processes (Fig. 4b,g) beautifully reproduces the fine structure of the experimental FT (Fig. 4a). Importantly, to capture the observation that scattering appears markedly unidirectional, a simple model taking into account a variation of the quasiparticle spectral weight associated with different bands is sufficient (Fig. 4c-f). In particular, the larger spectral weight of orbitals oriented along the vertical axis compared to the equivalent orbitals along the orthogonal direction explains the anisotropy in the experimental FT (Fig. 4a,g). Importantly, we note that anisotropic scattering from both in-plane ($d_{xy}/d_{x2-y2}$) and out-of-plane ($d_{xz}/d_{yz}$) $d$ orbitals is necessary to explain our observations, and as such all of them contribute to the electronic nematicity observed in this system.

Rotation symmetry breaking in V-based kagome superconductors has been reported to be CDW-driven [30] and intimately tied to the smectic CDW phase onset [34,35], which could in principle be explained by the inter-layer stacking order of the three-dimensional CDW phase [23,24]. In contrast to the smectic order in $AV_3Sb_5$, our experiments reveal the emergence of a pure electronic nematic state in $CsTi_3Bi_5$, in the absence of the CDW phase in cousin system $CsV_3Sb_5$. As such, our work shifts the paradigm of rotation symmetry breaking in kagome metals beyond any ties to CDW stacking and charge-stripe orders. Magnetotransport experiments further support the underlying rotation symmetry breaking in $CsTi_3Bi_5$ [46]. While no obvious anomalies are detected in zero-field resistivity, magnetization or heat capacity [37], it is possible that the signal may be too weak to be detected in such experiments.

There are several aspects of electronic nematicity in $CsTi_3Bi_5$ that stand out compared to other families of materials. First, in most other systems where evidence for electronic nematicity is observed, such as Cu and Fe-based superconductors, strong electronic correlation and proximity to magnetism make it is difficult to determine whether the nematic order is driven by spin or orbital degrees of freedom. The kagome metals have paramagnetic spin susceptibility without any magnetism, either localized or itinerant. It is therefore surprising that nematicity arises in $CsTi_3Bi_5$, which offers a fresh opportunity to explore an orbital driven nematic order. Second, electronic nematic order in hexagonal systems is a new subject with a potential for realizing three-state Potts nematic order [47]. The commonly studied materials, with the exception of twisted Moire structures [48–51], do not offer this possibility. However, in contrast to nematicity in Moire heterostructures where rotation symmetry breaking occurs upon a partial filling of localized electrons occupying flat bands, electronic nematicity in $CsTi_3Bi_5$ clearly affects delocalized Ti-derived $d$ orbitals. It is interesting to note that angular anisotropy of $\mathbf{q}_1$, connecting $p_z$ orbitals only, appears much smaller compared to $\mathbf{q}_2$ and $\mathbf{q}_3$ which involve $d$-orbitals (Supplementary Figure 10), further supporting the notion of electronic nematicity from $d$ electronic orbitals. How correlations of the electronic orbitals can lead to Potts nematicity is a new frontier of quantum materials, and our work provides ground work in that direction.

It is possible that the smectic CDW in $CsV_3Sb_5$ and the nematic order in $CsTi_3Bi_5$ share similar mechanisms, but the emergence of CDWs in $CsV_3Sb_5$ occurs due to stronger effective interactions, as correlation effects in $CsV_3Sb_5$ may be enhanced due to higher density of states in proximity to van Hove filling [52,53]. However we note that the two systems show fundamental differences, including the differing strengths of spin-orbit coupling and different energies of van Hove singularities. In contrast to $CsV_3Sb_5$ that has a pronounced lattice instability contributing to the emergence of the smectic CDW phase, no such instability exists in the phonon calculations of $CsTi_3Bi_5$ (Supplementary Figure 8). This points towards electronically-driven nematic state in $CsTi_3Bi_5$. A natural explanation could lie in orbital ordering, such as ferro-orbital or bond orbital ordering, where extended Coulomb interactions play a significant role [54]. The multi-orbital nature of $CsTi_3Bi_5$ with multiple Ti $d$ bands crossing the Fermi level is akin to that of Fe-based superconductors where several Fe $d$ bands span the Fermi level. It is conceivable that electronic nematicity uncovered here could be described by the nematic bond order proposed to explain the electronic nematic phase in FeSe [44,54] – but on a hexagonal kagome lattice distinct from the nearly square lattice of FeSe. Interestingly, if similarities between normal states of FeSe and $CsTi_3Bi_5$ extend to the superconducting regime, it would be of high interest to explore in future experiments if and how electronic nematicity in the normal state affects the mechanism of Cooper pairing in this family of kagome metals.

**Methods**

*Bulk single crystal synthesis and characterization.* Bulk single crystals of $CsBi_3Ti_5$ have been synthesized as described in Ref. [37].

*STM experiments.* Bulk single crystals of $CsTi_3Bi_5$ are transported from UCSB to BC in a sealed glass vile filled with argon gas to prevent air degradation. The vile is opened in an argon glove box at BC, where the sample is glued to a Unisoku-style STM sample holder, and a cleave bar is then attached to the sample. We quickly transfer the prepared sample from the glove box to an ultra-high vacuum load lock within seconds, and cold-cleave it before putting it into the microscope. STM data was acquired using a customized Unisoku USM1300 microscope at about 4.5 K. Spectroscopic measurements were performed using a standard lock-in technique with 910 Hz frequency and bias excitation noted in the figure captions. The STM tips used were home-made chemically etched tungsten tips, annealed in ultrahigh vacuum to bright orange color before STM experiments. To remove the effects of small piezoelectric and thermal drifts during the acquisition of data, we apply the Lawler–Fujita drift-correction algorithm on all our data, which aligns the atomic Bragg peaks in STM topographs to be equal in magnitude and oriented 60° apart.

*Ruling out STM tip artifacts.* To rule out the possibility of tip artifacts artificially inducing the electronic anisotropy, we characterize the electronic symmetry axis on the Bi surface, retract the tip, and manually rotate the sample 120 degrees. After the tip re-approaches on the sample, we take equivalent d$I$/d$V$ maps again. As shown, the electronic symmetry axis also rotates by 120° (Figure 3g,h). Furthermore, 4 different Bi regions are explored on 2 different CsTi$_3$Bi$_5$ samples, and all of them show the C$_2$-symmetric scattering and interference patterns (several examples shown in Supplementary Figure 3). In these regions, we manually bias-pulse the tip on the Bi surface (which changes the microscopic arrangement of atoms at the tip apex) to change the tip, and examine the data before and after the tip change. We demonstrate that similar anisotropic scattering and interference patterns can be resolved on the same region even with different tips (Supplementary Figure 3). As such, we deem that the C$_2$-symmetric electronic signature cannot be explained by STM tip artifacts.

**Density Functional Theory (DFT) calculation**. The structure of CsTi$_3$Bi$_5$ is fully relaxed with the Vienna *ab-initio* Simulation Package (VASP) [55], where the Perdew-Burke-Ernzerhof (PBE) [56] type generalized gradient approximation is employed to mimic the electron-electron interaction. The cutoff energy for the plane-wave basis set is 300 eV, and a *k*-mesh of 12*12*6 is used to sample the Brillouin zone. The DFT-D3 vdW correction [57] is considered in the relaxation. After that, the Full Potential Local Orbital (FPLO) software [58] is employed to calculate the band structure under PBE approximation with spin-orbital coupling considered. A *k* mesh size of 12*12*6 is used. The Fermi surface of CsTi$_3$Bi$_5$ is calculated from the tight-binding Hamiltonian as obtained from FPLO wannier fitting. The Ti *d* and Bi *p* orbitals are used as the wannier basis set.

*Note:* As we were finalizing our manuscript, we became aware of a similar experiment posted as a preprint [46].

## Acknowledgements

I.Z. gratefully acknowledges the support from NSF-DMR 2216080. S.D.W. and B.R.O. acknowledge support via the UC Santa Barbara NSF Quantum Foundry funded via the Q-AMASE-i program under award DMR-1906325. Z.W. acknowledges the support of U.S. Department of Energy, Basic Energy Sciences Grant No. DE-FG02-99ER45747 and the Cottrell SEED Award No. 27856 from Research Corporation for Science Advancement. D.W. and D.J. acknowledge support by the Bavaria California Technology Center BaCaTeC, Grant 7 [2021-2].

## Competing Interests

The Authors declare no Competing Financial or Non-Financial Interests.

## Code availability

The computer code used for data analysis is available upon request from the corresponding authors.

## Data Availability

The data supporting the findings of this study are available upon request from the corresponding authors.

**References**

1. Ortiz, B. R. *et al.* New kagome prototype materials: discovery of $KV_3Sb_5$, $RbV_3Sb_5$ and $CsV3Sb5$. *Phys. Rev. Mater.* **3**, 094407 (2019).

2. Ortiz, B. R. *et al.* $CsV_3Sb_5$: A $Z_2$ Topological Kagome Metal with a Superconducting Ground State. *Phys. Rev. Lett.* **125**, 247002 (2020).

3. Yin, Q. *et al.* Superconductivity and Normal-State Properties of Kagome Metal $RbV_3Sb_5$ Single Crystals. *Chinese Phys. Lett.* **38**, 037403 (2021).

4. Ortiz, B. R. *et al.* Superconductivity in the $Z_2$ kagome metal $KV_3Sb_5$. *Phys. Rev. Mater.* **5**, 034801 (2021).

5. Jiang, Y.-X. *et al.* Unconventional chiral charge order in kagome superconductor $KV_3Sb_5$. *Nat. Mater.* **20**, 1353–1357 (2021).

6. Zhao, H. *et al.* Cascade of correlated electron states in the kagome superconductor $CsV_3Sb_5$. *Nature* **599**, 216–221 (2021).

7. Li, H. *et al.* Rotation symmetry breaking in the normal state of a kagome superconductor $KV_3Sb_5$. *Nat. Phys.* **18**, 265–270 (2022).

8. Guo, C. *et al.* Switchable chiral transport in charge-ordered kagome metal $CsV_3Sb_5$. *Nature* **611**, 461–466 (2022).

9. Wu, Q. *et al.* Simultaneous formation of two-fold rotation symmetry with charge order in the kagome superconductor $CsV_3Sb_5$ by optical polarization rotation measurement. *Phys. Rev. B* **106**, 205109 (2022).

10. Mielke, C. *et al.* Time-reversal symmetry-breaking charge order in a kagome superconductor. *Nature* **602**, 245–250 (2022).

11. Chen, H. *et al.* Roton pair density wave in a strong-coupling kagome superconductor. *Nature* **599**, 222–228 (2021).

12. Liang, Z. *et al.* Three-Dimensional Charge Density Wave and Surface-Dependent Vortex-Core States in a Kagome Superconductor $CsV_3Sb_5$. *Phys. Rev. X* **11**, 031026 (2021).

13. Li, H. *et al.* Observation of Unconventional Charge Density Wave without Acoustic Phonon Anomaly in Kagome Superconductors $AV_3Sb_5$ (A=Rb, Cs). *Phys. Rev. X* **11**, 031050 (2021).

14. Kang, M. *et al.* Twofold van Hove singularity and origin of charge order in topological kagome superconductor $CsV_3Sb_5$. *Nat. Phys.* **18**, 301–308 (2022).

15. Xiang, Y. *et al.* Twofold symmetry of c-axis resistivity in topological kagome superconductor $CsV_3Sb_5$ with in-plane rotating magnetic field. *Nat. Commun.* **12**, 6727 (2021).

16. Yang, S.-Y. *et al.* Giant, unconventional anomalous Hall effect in the metallic frustrated magnet candidate, $KV_3Sb_5$. *Sci. Adv.* **6**, eabb6003 (2020).

17. Xu, H.-S. *et al.* Multiband Superconductivity with Sign-Preserving Order Parameter in Kagome Superconductor $CsV_3Sb_5$. *Phys. Rev. Lett.* **127**, 187004 (2021).

18. Uykur, E., Ortiz, B. R., Wilson, S. D., Dressel, M. & Tsirlin, A. A. Optical detection of the density-wave instability in the kagome metal $KV_3Sb_5$. *npj Quantum Mater.* **7**, 16 (2022).

19. Feng, X., Jiang, K., Wang, Z. & Hu, J. Chiral flux phase in the Kagome superconductor $AV_3Sb_5$.

*Sci. Bull.* **66**, 1384–1388 (2021).

20. Tan, H., Liu, Y., Wang, Z. & Yan, B. Charge Density Waves and Electronic Properties of Superconducting Kagome Metals. *Phys. Rev. Lett.* **127**, 046401 (2021).

21. Denner, M. M., Thomale, R. & Neupert, T. Analysis of Charge Order in the Kagome Metal $AV_3Sb_5$ (A=K, Rb, Cs). *Phys. Rev. Lett.* **127**, 217601 (2021).

22. Lin, Y.-P. & Nandkishore, R. M. Complex charge density waves at Van Hove singularity on hexagonal lattices: Haldane-model phase diagram and potential realization in the kagome metals $AV_3Sb_5$ (A=K, Rb, Cs). *Phys. Rev. B* **104**, 045122 (2021).

23. Park, T., Ye, M. & Balents, L. Electronic instabilities of kagome metals: Saddle points and Landau theory. *Phys. Rev. B* **104**, 035142 (2021).

24. Christensen, M. H., Birol, T., Andersen, B. M. & Fernandes, R. M. Theory of the charge density wave in $AV_3Sb_5$ kagome metals. *Phys. Rev. B* **104**, 214513 (2021).

25. Zhou, S. & Wang, Z. Chern Fermi pocket, topological pair density wave, and charge-4e and charge-6e superconductivity in kagomé superconductors. *Nat. Commun.* **13**, 7288 (2022).

26. Christensen, M. H., Birol, T., Andersen, B. M. & Fernandes, R. M. Loop currents in $AV_3Sb_5$ kagome metals: Multipolar and toroidal magnetic orders. *Phys. Rev. B* **106**, 144504 (2022).

27. Liu, Z. Z. *et al.* Charge-Density-Wave-Induced Bands Renormalization and Energy Gaps in a Kagome Superconductor $RbV_3Sb_5$. *Phys. Rev. X* **11**, 041010 (2021).

28. Hu, Y. *et al.* Topological surface states and flat bands in the kagome superconductor $CsV_3Sb_5$. *Sci. Bull.* **67**, 495–500 (2022).

29. Luo, H. *et al.* Electronic nature of charge density wave and electron-phonon coupling in kagome superconductor $KV_3Sb_5$. *Nat. Commun.* **13**, 273 (2022).

30. Nie, L. *et al.* Charge-density-wave-driven electronic nematicity in a kagome superconductor. *Nature* **604**, 59–64 (2022).

31. Ni, S. *et al.* Anisotropic Superconducting Properties of Kagome Metal $CsV_3Sb_5$. *Chinese Phys. Lett.* **38**, 057403 (2021).

32. Zhou, X. *et al.* Origin of charge density wave in the kagome metal $CsV_3Sb_5$ as revealed by optical spectroscopy. *Phys. Rev. B* **104**, L041101 (2021).

33. Oey, Y. M. *et al.* Fermi level tuning and double-dome superconductivity in the kagome metal $CsV_3Sb_{5-x}Sn_x$. *Phys. Rev. Mater.* **6**, L041801 (2022).

34. Xu, Y. *et al.* Three-state nematicity and magneto-optical Kerr effect in the charge density waves in kagome superconductors. *Nat. Phys.* **18**, 1470–1475 (2022).

35. Li, H. *et al.* Unidirectional coherent quasiparticles in the high-temperature rotational symmetry broken phase of $AV_3Sb_5$ kagome superconductors. *Nat. Phys.* **19**, 637–643 (2023).

36. Yang, H. *et al.* Titanium-based kagome superconductor $CsTi_3Bi_5$ and topological states. *ArXiv* 2209.03840 (2022).

37. Werhahn, D. *et al.* The kagomé metals $RbTi_3Bi_5$ and $CsTi_3Bi_5$. *Zeitschrift für Naturforsch. B* **77**, 757–764 (2022).

38. Liu, B. *et al.* Tunable van Hove singularity without structural instability in Kagome metal

CsTi$_3$Bi$_5$. *ArXiv* 2212.04460 (2022).

39. Rosenthal, E. P. *et al.* Visualization of electron nematicity and unidirectional antiferroic fluctuations at high temperatures in NaFeAs. *Nat. Phys.* **10**, 225–232 (2014).

40. Chuang, T.-M. *et al.* Nematic electronic structure in the 'parent' state of the iron-based superconductor Ca(Fe$_{1-x}$Co$_x$)$_2$As$_2$. *Science* **327**, 181–4 (2010).

41. Kostin, A. *et al.* Imaging orbital-selective quasiparticles in the Hund's metal state of FeSe. *Nat. Mater.* **17**, 869–874 (2018).

42. Singh, U. R. *et al.* Evidence for orbital order and its relation to superconductivity in FeSe$_{0.4}$Te$_{0.6}$. *Sci. Adv.* **1**, e1500206 (2015).

43. Zhao, H. *et al.* Nematic transition and nanoscale suppression of superconductivity in Fe(Te,Se). *Nat. Phys.* **17**, 903–908 (2021).

44. Zhang, P. *et al.* Observation of two distinct dxz/dyz band splittings in FeSe. *Phys. Rev. B* **91**, 214503 (2015).

45. Wang, Q. *et al.* Charge Density Wave Orders and Enhanced Superconductivity under Pressure in the Kagome Metal CsV$_3$Sb$_5$. *Adv. Mater.* **33**, 2102813 (2021).

46. Yang, H. *et al.* Superconductivity and orbital-selective nematic order in a new titanium-based kagome metal CsTi$_3$Bi$_5$. *ArXiv* 2211.12264 (2022).

47. Fernandes, R. M. & Venderbos, J. W. F. Nematicity with a twist: Rotational symmetry breaking in a moiré superlattice. *Sci. Adv.* **6**, eaba8834 (2020).

48. Kerelsky, A. *et al.* Maximized electron interactions at the magic angle in twisted bilayer graphene. *Nature* **572**, 95–100 (2019).

49. Jiang, Y. *et al.* Charge order and broken rotational symmetry in magic-angle twisted bilayer graphene. *Nature* **573**, 91–95 (2019).

50. Xie, Y. *et al.* Spectroscopic signatures of many-body correlations in magic-angle twisted bilayer graphene. *Nature* **572**, 101–105 (2019).

51. Choi, Y. *et al.* Electronic correlations in twisted bilayer graphene near the magic angle. *Nat. Phys.* **15**, 1174–1180 (2019).

52. Kivelson, S. A., Fradkin, E. & Emery, V. J. Electronic liquid-crystal phases of a doped Mott insulator. *Nature* **393**, 550–553 (1998).

53. Wang, W.-S., Li, Z.-Z., Xiang, Y.-Y. & Wang, Q.-H. Competing electronic orders on kagome lattices at van Hove filling. *Phys. Rev. B* **87**, 115135 (2013).

54. Jiang, K., Hu, J., Ding, H. & Wang, Z. Interatomic Coulomb interaction and electron nematic bond order in FeSe. *Phys. Rev. B* **93**, 115138 (2016).

55. Kresse, G. & Furthmüller, J. Efficiency of ab-initio total energy calculations for metals and semiconductors using a plane-wave basis set. *Comput. Mater. Sci.* **6**, 15–50 (1996).

56. Perdew, J., Burke, K. & Ernzerhof, M. Generalized Gradient Approximation Made Simple. *Phys. Rev. Lett.* **77**, 3865–3868 (1996).

57. Grimme, S., Antony, J., Ehrlich, S. & Krieg, H. A consistent and accurate ab initio parametrization of density functional dispersion correction (DFT-D) for the 94 elements H-Pu. *J. Chem. Phys.* **132**,

154104 (2010).

58. Koepernik, K. & Eschrig, H. Full-potential nonorthogonal local-orbital minimum-basis band-structure scheme. *Phys. Rev. B* **59**, 1743–1757 (1999).

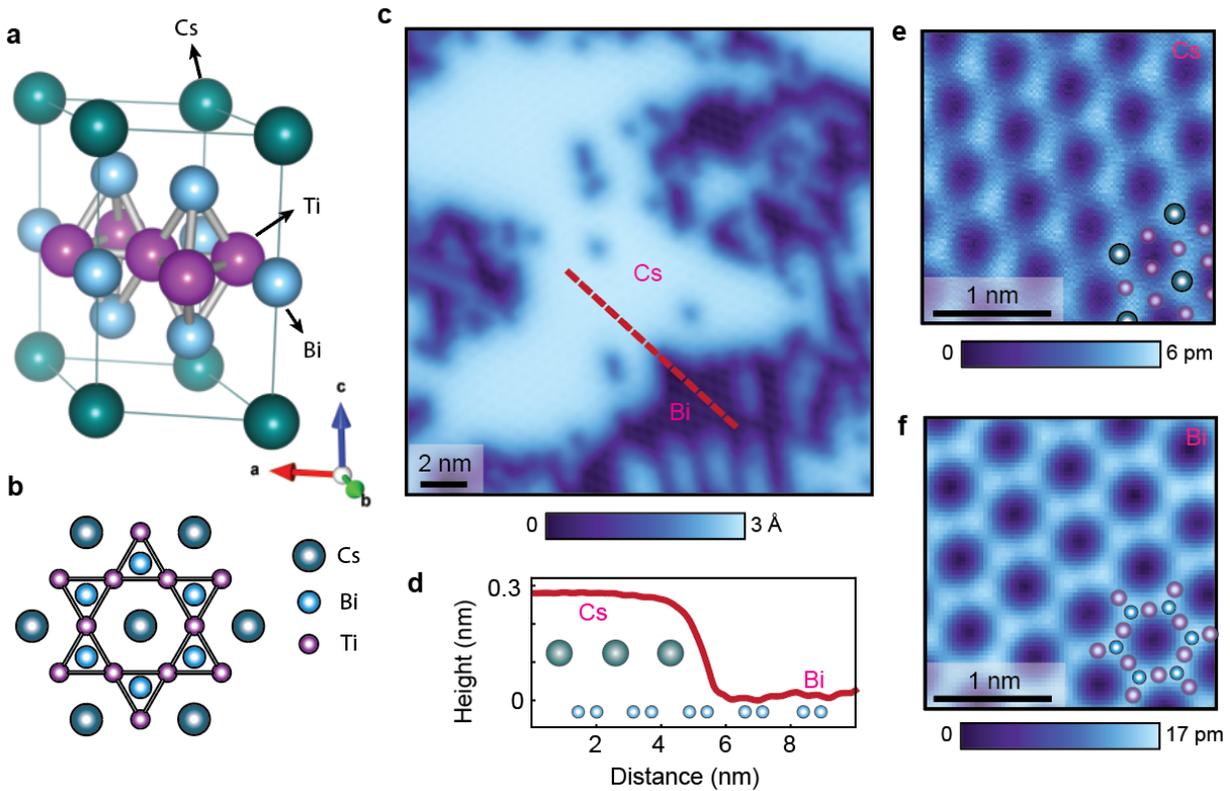

**Figure 1. Crystal structure and topographic characterization of CsTi$_3$Bi$_5$.** **(a)** A 3D ball model of the crystal structure of CsTi$_3$Bi$_5$. **(b)** A 2D ball model showing relative positions of different atoms in the *ab*-plane. **(c)** A 20 nm square scanning tunneling microscopy (STM) topograph of CsTi$_3$Bi$_5$ taken at 4.5 K showing Cs islands on top of a complete Bi termination. **(d)** A topographic line profile taken along the red dashed line in (c) across a step between the two terminations. **(e,f)** Zoomed-in STM topographs of smaller regions of the Cs (e) and the Bi (f) surface (see also Supplementary Figure 6). Bottom right corner insets in (e,f) show a superimposed atomic structure. STM setup conditions: $I_{set}$ = 200 pA, $V_{sample}$ = 100 mV (c); $I_{set}$ = 400 pA, $V_{sample}$ = 100 mV (e); $I_{set}$ = 300 pA, $V_{sample}$ = 100 mV (f).

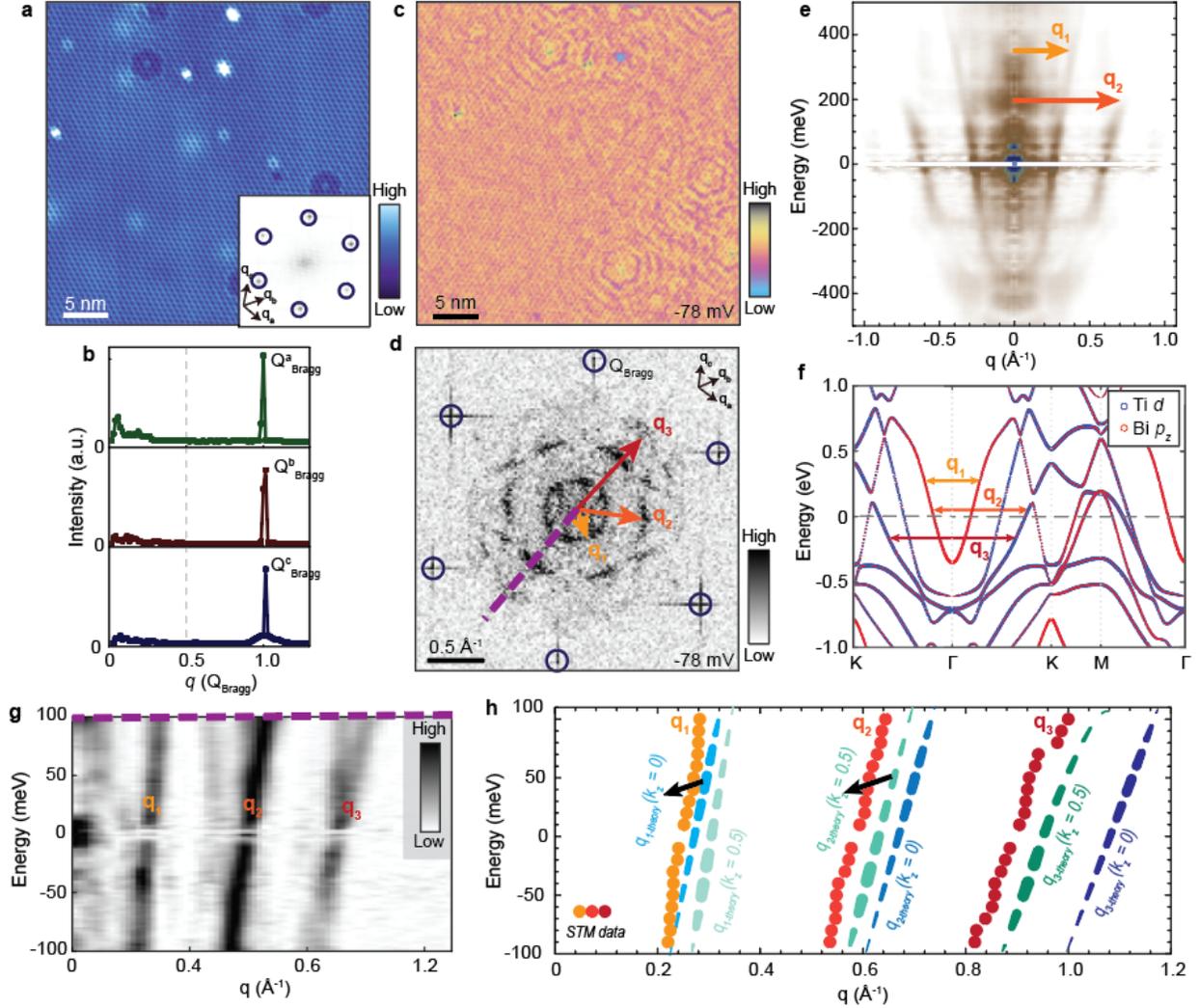

**Figure 2. Large-scale electronic band structure and the absence of charge density wave in $CsTi_3Bi_5$.**
**(a)** STM topograph taken over a clean Bi-terminated region. The bottom right inset is its Fourier transform (FT) showing a hexagonal lattice. **(b)** FT linecuts starting from the center of the FT in the inset in (a) along the three atomic Bragg peak directions, showing the absence of the 2 x 2 charge density wave peaks present in cousin $CsV_3Sb_5$. **(c)** An example of a normalized conductance map $L(\mathbf{r},V)=((dI(\mathbf{r},V)/dV)/(I(\mathbf{r},V)/V))$ taken approximately over the region in (a), and **(d)** its FT. Yellow, orange and red arrows denote the three dominant scattering wave vectors $\mathbf{q_1}$, $\mathbf{q_2}$ and $\mathbf{q_3}$, respectively. **(e)** Radially averaged linecut extracted from the FTs of $L(\mathbf{r},V)$ maps. The data near 0 mV is artificially suppressed due to the divergent behavior of $L(\mathbf{r},V)$ maps for near-zero bias $V$. **(f)** Density functional theory (DFT) calculated band structure of $CsTi_3Bi_5$ along high-symmetry directions K-Γ-K-M-Γ, with different colors denoting either Bi (red) or Ti (blue) orbital character. **(g)** A zoomed-in FT linecut near Fermi level taken along the purple dashed line in (d) across the three wave vectors. **(h)** Comparisons between dispersions of $\mathbf{q_i}$ (i = 1,2,3) and equivalent vectors extracted from the DFT band structure for $k_z = 0$ (blue) and $k_z = 0.5$ (green lines). STM data points are determined by fitting a Gaussian peak function to the linecut in (g) at energies separated by 10 meV increments. STM setup conditions: $I_{set}$ = 200 pA, $V_{sample}$ = 100 mV (a); $I_{set}$ = 600 pA, $V_{sample}$ = 100 mV, $V_{exc}$ = 4 mV (c,d,g); $I_{set}$ = 1 nA, $V_{sample}$ = 500 mV, $V_{exc}$ = 10 mV (e). Data were taken at 4.5 K.

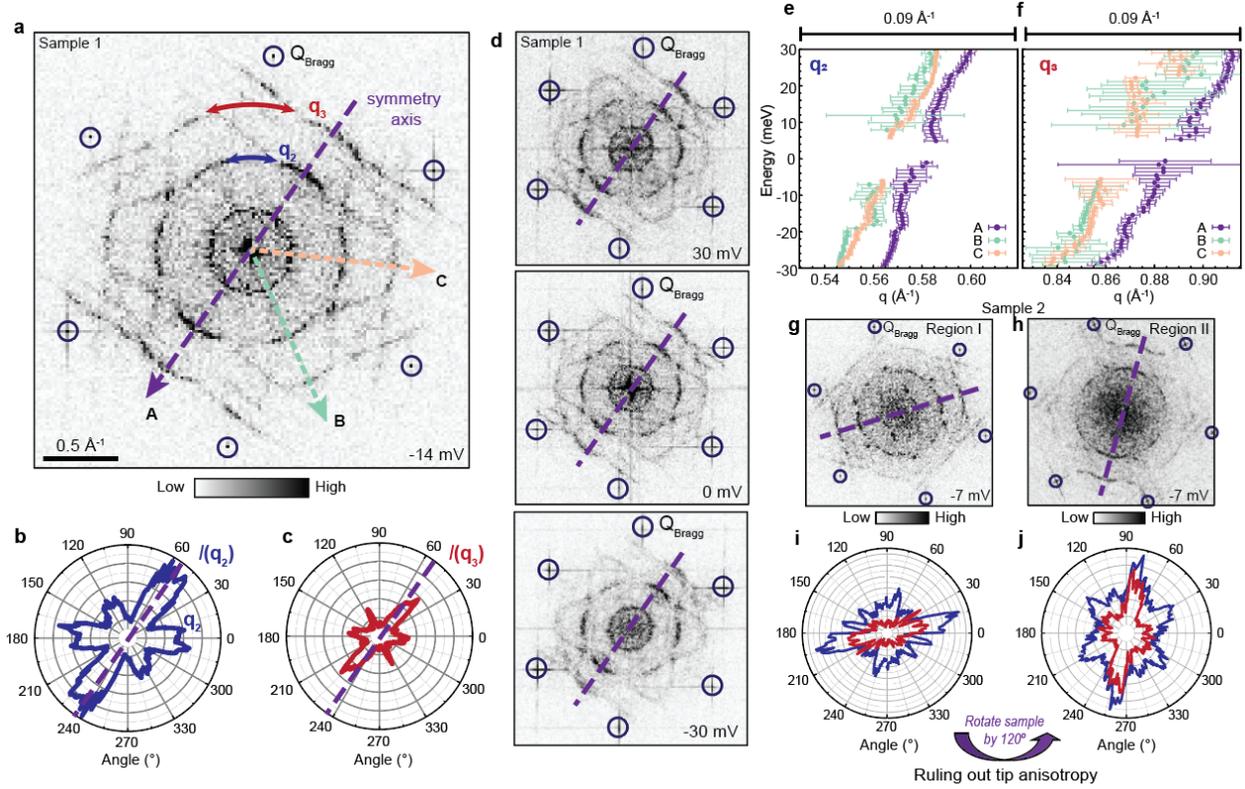

**Figure 3. Two-fold symmetric electronic signature. (a)** Fourier transform of a normalized differential conductance $L(\mathbf{r}, V=-14\text{ mV})$ map taken on the Bi surface. Atomic Bragg peaks are circled in black. Blue and red double-arrows depict where $\mathbf{q}_2$ and $\mathbf{q}_3$ wave vector amplitudes are extracted. **(b,c)** Angle-dependent Fourier transform (FT) amplitudes of $\mathbf{q}_2$ and $\mathbf{q}_3$ plotted in polar coordinates on the same scale. The purple dashed lines in (a-c) denote the apparent $C_2$ symmetry axis. **(d)** Representative energy-dependent FTs also showing the two-fold symmetric scattering signature. **(e,f)** Momentum-transfer space positions of the wave vectors $\mathbf{q}_2$ and $\mathbf{q}_3$ along the three Γ-K directions. As it can be seen, both $\mathbf{q}_2$ and $\mathbf{q}_3$ are slightly larger along direction A, compared to the other two directions B and C, which appear nearly indistinguishable. Directions A, B and C are defined in panel (a). Error bars in (e,f) represent the standard errors obtained by Lorentzian fits to the data. **(g,h)** FTs of the $L(\mathbf{r}, V)$ maps taken over two different regions on the same $CsTi_3Bi_5$ sample #2, denoted as regions I and II, both of the Bi surface. Region II is found after a manual rotation of the whole sample by 120 degrees, while the tip wire in both cases remains the same. Atomic Bragg peaks are circled in black. **(i,j)** Angle-dependent FT amplitudes of $\mathbf{q}_2$ (blue) and $\mathbf{q}_3$ (red line) plotted in polar coordinates for data in (g,h). The symmetry axis is approximately along the 15° angle in (i), and along the 255° angle in (j). STM setup conditions: $I_{set} = 300$ pA, $V_{sample} = 30$ mV, $V_{exc} = 4$ mV (a,d); $I_{set} = 300$ pA, $V_{sample} = 30$ mV, $V_{exc} = 4$ mV (e-f); $I_{set} = 200$ pA, $V_{sample} = -10$ mV, $V_{exc} = 2$ mV (g); $I_{set} = 500$ pA, $V_{sample} = 10$ mV, $V_{exc} = 5$ mV (h).

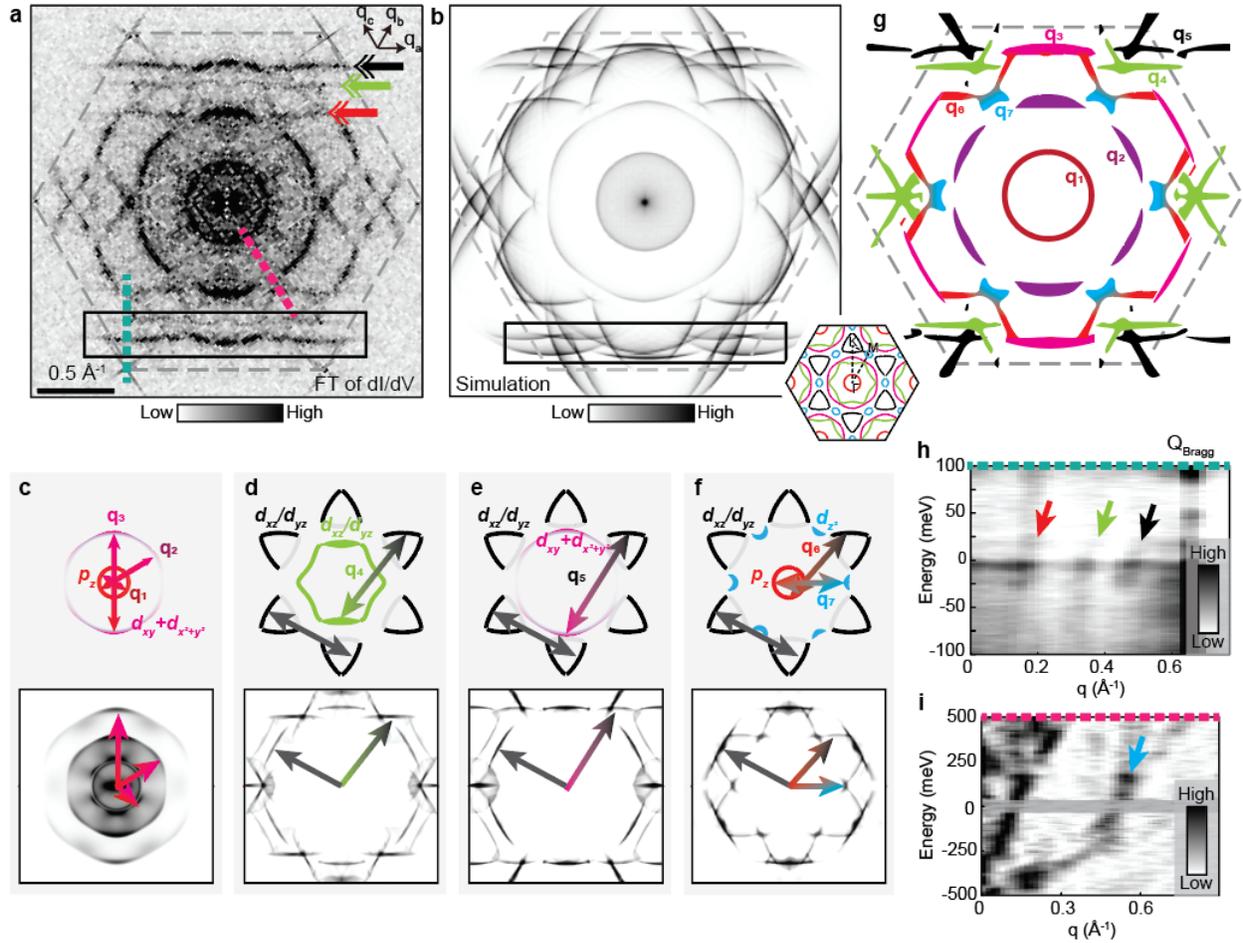

**Figure 4. Mapping the origin of quasiparticle scattering and interference. (a)** Two-fold symmetrized Fourier transform (FT) of the $L(\mathbf{r}, V=-14\text{ mV})$ map also shown in Fig. 3a, rotated so that the symmetry axis is along the $x$-axis. Vertices of the large gray dashed hexagon connect the six atomic Bragg peaks. **(b)** A joint density of states (JDOS) simulation at 0 meV for $k_z = 0.5$, only considering the scattering processes $\mathbf{q_1}$ to $\mathbf{q_7}$ shown in (c-f) to emphasize the relevant features. No spectral weight anisotropy is considered in (b). As it can be seen by visual comparison, the morphology of main scattering wave vectors can be captured to a great detail. Bottom right inset in (b) is a complete CEC at zero energy from our DFT calculations at $k_z=0.5$. For visual purposes, different colors in the CEC are chosen to match the colors of bands in (c-f). The largest hexagon in the inset of (b) represents the outline of the second Brillouin zone. **(c-f)** Schematic of the constant energy contour (CEC) at Fermi energy for $k_z=0.5$ with only a subset of relevant bands (top panel) and its mathematical auto-correlation (bottom panel). Different thicknesses of lines in each CEC represent a varying spectral weight of the associated portion of each band. **(g)** A schematic of main scattering wave vectors obtained from our analysis in (c-f) put together, taking into account anisotropic spectral weight, for a direct comparison with data in (a). Each part of the schematic is color-coded to match the **q** vectors in (a, c-f) to more easily visually portray the scattering origin. **(h,i)** FT linecuts as a function of energy along the $q$-space lines denoted by dashed green (h) and dashed pink (i) lines in (a). The dispersive nature of these features is apparent.